\documentclass[prd,aps,floats,twocolumn,nofootinbib]{revtex4-1}
\usepackage{slashed}
\usepackage{mathtools}
\usepackage{amsfonts}
\usepackage{amssymb}
\usepackage{epsfig}
\usepackage{rotating}
\usepackage{url}
\usepackage{times}
\usepackage{color}
\usepackage{bm}
\usepackage{xcolor,colortbl}
\usepackage{hyperref}
\usepackage[alsoload=hep]{siunitx}
\usepackage{enumitem}
\usepackage{fancyhdr}
\usepackage{anyfontsize}
\usepackage{wasysym}
\usepackage{empheq}

\newcommand{\nn}{\nonumber}
\newcommand{\ph}{\phantom}
\newcommand{\eps}{\epsilon}

\newcommand{\be}{\begin{equation}}
\newcommand{\ee}{\end{equation}}
\newcommand{\bea}{\begin{eqnarray}}
\newcommand{\eea}{\end{eqnarray}}

\newcommand{\plk}{\mathfrak{h}}

\begin{document}

%%%%%%%%%%%%%%%%%%%%%%%%%%%%%

\title{Quantum torsion and a Hartle-Hawking ``beam''}
\author{Jo\~{a}o Magueijo}
\email{j.magueijo@imperial.ac.uk}
\affiliation{Theoretical Physics Group, The Blackett Laboratory, Imperial College, Prince Consort Rd., London, SW7 2BZ, United Kingdom}
\author{Tom Zlosnik}
\email{zlosnik@fzu.cz}
\affiliation{CEICO, Institute of Physics of the Czech Academy of Sciences, Na Slovance 1999/2, 182 21, Prague}
\date{\today}

\begin{abstract}
In the Einstein-Cartan framework the torsion-free conditions arise within the Hamiltonian treatment as second-class constraints.
The standard strategy is to solve these constraints, eliminating the torsion from the classical theory, before quantization. Here we advocate 
leaving the torsion inside the other constraints before quantization, leading at first to wave functions that can be called ``kinematical'' with regards to
the torsion, but not the other constraints. The torsion-free condition can then be imposed as a condition upon the physical wave 
packets one constructs, satisfying the usual uncertainty relations, and so with room for quantum fluctuations in the torsion. 
This alternative strategy has the surprising effect of clarifying the sense in which the wave functions solving an explicitly real theory are ``delta-function normalizable''. Such solutions with zero (or any fixed) torsion, should be interpreted as plane waves in torsion space. Properly constructed wave packets 
are therefore normalizable in the standard sense. Given that they are canonical duals, this statement applies equally well to the Chern-Simons state (connection 
representation) and the Hartle-Hawking wave function (metric representation). 
We show how, when torsion is taken into account, the Hartle-Hawking wave function is replaced by a Gauss-Airy function, with finite norm,
which we call the Hartle-Hawking beam. The Chern-Simons
state, instead, becomes a packet with a  Gaussian probability distribution in connection space. We conclude the paper with two sections 
explaining how to generalize  these results beyond minisuperspace. 
\end{abstract}

\maketitle

\section{Introduction}
Classical general relativity may be variously formulated giving primacy to the metric (the second order formalism), or 
allowing the metric and the connection to be initially independent variables (the first order formalism). The latter may be 
motivated in many different ways (e.g.~\cite{Kibble,Hehl}), but historically has its roots in the Einstein-Cartan formalism. 
In the minimal theories, in the absence of spinors, the first and second order formulations are classically equivalent, since
one of the Einstein-Cartan equations of motion relates the torsion to the spin current (which we will assume vanishes throughout this paper). 
A question arises concerning the quantum mechanics of the two formulations: are they equivalent? 
In the Einstein-Cartan formalism zero-torsion appears classically as an equation of motion 
(and not a built-in fact, as is the case in the metric formalism, with a Christoffel connection). 
Quantum mechanics probes the phase space off-shell, i.e. away from the classical solutions,  so it could feel the non-zero torsion {\it possibility}
of Einstein-Cartan theory. Could quantum torsion fluctuations lead to an inequivalent quantum theory?

A closer inspection reveals that the issue is technically non-trivial. In the canonical framework (used in most quantization schemes), the torsion-free
condition appears as part of the secondary constraints, which turn out to be second class (e.g.~\cite{notimegauge,realK}). 
Quantization using the usual Poisson bracket
is therefore inconsistent. One approach to second class constraints is to employ the Dirac bracket in lieu of the Poisson bracket~\cite{Diracbk}.
Another approach is, if at all possible, to solve the constraints classically, inserting the resulting equations
of motion into the action before quantizing. This is usually the route taken regarding the torsion. 
But is this right? Is this the only path to quantization? It certainly imposes a freeze on torsion quantum fluctuations. This, we shall argue in this paper, is the origin of some apparent infinite norms found in the usual approach.

We have in mind the fact that some well-known solutions in quantum gravity are not normalizeable in the conventional sense. 
This can be seen both in the metric and connection representation, since the norm is independent of the  representation. 
In the connection representation, the matter concerns the Chern-Simons wave function 
(also called the Kodama state~\cite{jackiw,witten,kodama,lee1,lee2}). Here, we must distinguish two types of normalization problems. If we take
the state that solves the Lorentzian complex Hamiltonian equation, then is suffers from a number of acknowledged
pathologies (e.g.~\cite{witten}):
CPT violating properties (and consequent impossibility of a positive energy property), lack of gauge invariance under large 
gauge transformations, and, indeed, non-normalizability with respect to the standard inner product.
However, if one employs an explicitly real theory (phase space variables and action) these problems disappear, as shown in~\cite{realK,MSS}. 
Then, the ``real'' Chern-Simons state is a pure phase in the Lorentzian signature (just like in the Euclidean signature), and so it 
is ``delta-function'' normalizable~\cite{Randono0,Randono1,Randono2}. It is in this last format that we want to examine the normalizability of the 
Chern-Simons state. 

But foremost, we want to highlight the fact that this is nothing but the dual of a seemingly unrelated problem. 
The Hartle-Hawking (HH) wave function is the Fourier 
dual of the Chern-Simons state reduced to mini-superspace, assuming real variables throughout~\cite{CSHHV}.
Its normalizability issues are well-known. We may take the view that one needs a higher dimensional space to measure and probability, from a Klein-Gordon current, as suggested by  Vilenkin~\cite{vil0,Vilenkin}. But we may also take the problem face value and wonder what it means. Flipping the question to the 
Fourier dual, the issue is the same as the one for the Chern-Simons state, and the question becomes: what does it mean to say that the Hartle-Hawking/Chern-Simons wave functions
are ``delta-function'' normalizable? With respect to what?

``Quantum torsion'' provides an answer, as we show in this paper. We start by reviewing the mini-superspace structures behind the equivalence of the Chern-Simons and Hartle-Hawking states:
complementary variables, Fourier transform  and integration measure (Section~\ref{background}). We also explain how the torsion is usually forced to be 
zero by means of second class constraints. We then show how solutions to the quantum constraints may be obtained by first ignoring the torsion conditions 
(Section~\ref{CSHHT}). This leads to torsionful versions of the Hartle-Hawking and Chern-Simons states. Computing the norms of these states, at once we see why the result
is infinite if the torsion is fixed (at zero or otherwise), and proportional to a delta-function in terms of the torsion (Section~\ref{norms}). 

With this important piece of information in hand we may now build wave packets in torsion space, which are normalizable in the conventional sense
(Section~\ref{wavepacks}). We may also build coherent states centred at zero. These are the physical states with regards to the torsion conditions.
We build these states explicitly in both representations (Section~\ref{secHHbeam}). The uniform probability extended over an infinite domain, implied
by the unbridled Chern-Simons state, is replaced by a Gaussian distribution. The Hartle-Hawking wave function is replaced by a regularized Gauss-Airy function: the 
Hartle-Hawking beam. Suddenly it all makes sense. 

We close with two sections explaining how to extend our construction beyond mini-superspace. They do not impact on our main results, but may
be an important first step towards the phenomenology of quantum torsion. 

%(or any other value, should
%we want to extend our results to quasi-topological theories~\cite{alex1,alex2,realK}. 

%Fixing the torsion to zero is behind the infinite norm. The amplitude is a delta function. The integral of the square of a delta function
%is infinity. [SHIFT TO THE END?] 

%Let us separate different types of torsion. [shift to end?]

\section{Review of mini-superspace structures}\label{background}

The Einstein-Cartan action reduced to homogeneity and isotropy is~\cite{CSHHV,MZ}:
\begin{equation}\label{Sg}
S=6\kappa V_{c}\int dt\bigg(a^{2}\dot{b}+Na\bigg(b^{2}+k-c^2 -\frac{\Lambda }{3}%
a^{2}\bigg)\bigg).
\end{equation}
where $\kappa=1/(16\pi G_N)$, $k=0,\pm1$ is the normalized spatial curvature, and $\Lambda$ is the cosmological constant. Here $a$ is the expansion factor
(the only metric variable), and $b$ and $c$ are components of the connection, respectively the
off-shell version of the Hubble parameter (since $b\approx \dot a$ if there is not torsion; see below), and the parity-violating
Cartan spiral staircase~\cite{spiral,MZ}. The Lagrange multiplier $N$ is the lapse function. 

Hence, the Poisson bracket is:
\begin{equation}\label{PB1}
\{b,a^{2}\}=\frac{1}{6\kappa V_{c}},
\end{equation}%
%and since the constraints are always first class, this implies commutation relations:
inducing the mini-superspace commutator: 
\begin{equation} \label{com1}
\left[ \hat{b},\hat{a^{2}}\right] =\frac{il_{P}^{2}}{3V_{c}},
\end{equation}%
where $l_{P}=\sqrt{8\pi G_{N}\hbar }$ is the reduced Planck length. Given that it will appear recurrently, to simplify the 
notation we define:
\be
\plk=\frac{l_{P}^{2}}{3V_{c}}.
\ee
We may also see $\plk$ as an ``effective'' Planck constant~\cite{Barrowhbar}, but we do not need to accept this interpretation.

The canonical structure and commutator imply the Fourier transform and its inverse:
\bea
\psi_{a^2}(a^2)&=&\int  \frac{db }{\sqrt{2\pi\plk }} e^{-\frac{i}{\plk}a^2 b}\psi_b(b),\nn\\
\psi_b(b)&=&\int \frac{da^2}{\sqrt{2\pi\plk }} e^{\frac{i}{\plk}a^2 b}\psi_a(a^2). \label{FT}
\eea
for changing between duals (where we have used a symmetric definition for 
FT and its inverse). The associated inner product between states and integration measure
are the trivial:
\bea
\langle \psi |\phi \rangle&=& \int
%\frac{db}{ \sqrt{2\pi}\plk }
db \,\psi^\star _{b}(b) \phi_{b}(b)\label{bmeas}\\
&=&\int
%\frac{da^2}{ \sqrt{2\pi}\plk }
da^2\, \psi^\star _{a^2}(a^2) \phi_{a^2}(a^2).\label{ameas}
\eea
All integrations will be understood to be over the whole real line here and from now on; we will mention alternatives at the end of the paper.
The Fourier transform can then be seen as an insertion of a partition of identity with:
\bea
\psi_{a^2}(a^2)&=&\langle a^2|\psi\rangle\label{a2rep}\\
 \psi_{b}(b)&=&\langle b|\psi\rangle\label{brep}
\eea
and
\be
\langle b|a^2\rangle=\frac{e^{\frac{i}{\plk}a^2 b}}{\sqrt{2\pi \plk}}
\ee

As explained in~\cite{CSHHV} (ignoring torsion), the solution to the Hamiltonian constraint in the $a^2$ representation is the Hartle-Hawking wave function, whereas in the $b$ representation it is the Chern-Simons state. The two are the Fourier transform of each other (with the definitions of measure and Fourier transform induced by the canonical structure, which we have just presented). The
theory's Hamiltonian has the form $H=  6\kappa V_c Na {\cal H}$ with:
\bea
{\cal H}&=&-b^{2}-k+c^2 +\frac{\Lambda }{3}%.
a^{2}.
\eea
But what about the conditions forcing the torsion to vanish classically?

%\subsection{Canonical treatment of the torsion}
In the canonical framework we should distinguish between two different types of torsion. 
One type of torsion is that contained in $b$, the mini-superspace reduction of the extrinsic curvature (the space-time components
of the spin-connection~\cite{alex2,MZ}, also denoted $K^i$ in the Ashtekar formalism~\cite{realK}). Specifically:
\be
b=\dot a/N-T
\ee
the first term containing the torsion-free component, the second the torsion. The Hamilton equation for $a$ reads:
\be
\dot a=Nb
\ee
so that it fixes the torsion $T$ to zero on-shell, {\it as part of the time evolution}. 

A quite different type of torsion is that contained in $c$, the mini-superspace version of the torsion contained in the purely spatial components 
of the spin-connection (denoted $\Gamma^i$ in the Ashtekar formalism~\cite{realK}). 
In mini-superspace (and more generally in the time gauge) $c$ (the torsion in $\Gamma^i$) does not have a momentum. This can be phrased by adding to the action a term corresponding to the Legendre transform between $c$ and its conjugate momentum $p_c$, plus a constraint forcing $p_c$ to be zero:
\be\label{Sc}
%S\rightarrow S-  6\kappa V_c \int dt (\dot c p_c - \lambda  p_c)
S\rightarrow S+ \int dt (\dot c p_c - \lambda  p_c)
\ee
%(the pre-factor merely a matter of later convenience; but see~\cite{realK}). 
Then,
\be\label{cpcPB}
%\{p_c,c\}=\frac{1}{6\kappa V_{c}}
\{c,p_c\}=1
\ee
but since:
\be
%\{p_c,{\cal H}\}=-\frac{c}{3\kappa V_{c}}
\{p_c,{\cal H}\}= 2 c
\ee
we get the secondary constraint:
\be
c\approx 0.
\ee
However because of (\ref{cpcPB}), the two new constraints ($c\approx 0$ and $p_c\approx 0$)
are second class constraints.

%How to deal with the torsion?
The usual argument is that second class constraints should be solved classically, before quantization, because they cannot be imposed consistently 
at the quantum mechanical level\footnote{Certainly we cannot impose
${\hat c}\psi={\hat p_c}\psi=0$, in whatever rep. Note that if we imposed only $p_c\approx 0$ that would be fine, and only 
state that the wave function could not depend on $c$, forcing a uniform distribution in $c$. Would this be inconsistent with ``observation''?}.
Thus one sets $c=0$ in the Hamiltonian equation, quantizes, and gets on with it. 
But is this the only way to quantize? Certainly not (we recall Gupta Bleuler quantization). 
%[for example reality conditions go into the inner product and the definition of physical states. Gupta Bleuler. Etc etc.].  
Here we propose an alternative, which happens to explain what is meant by delta-function normalization of the Chern-Simons and Hartle-Hawking wave functions. 

\section{Kinematical torsionful wave functions}\label{CSHHT}

Let us solve the quantum Hamiltonian constraint, leaving the $c$ there, to be seen as a parameter, just like the 
(torsion-free) spatial curvature $k=0,\pm 1$. 
%Alternatively, we can consider $c$ to be an observable and state that we are working in a representation diagonalizing $c$. 
As in~\cite{CSHHV} we find that with suitable ordering, 
the solution to the Hamiltonian constraint in the $b$ representation is the torsionful version of the Chern-Simons state
in mini-superspace:
\be\label{kodbc1}
\psi_{CS }(b,c)= {\cal N} \exp{\bigg(i\frac{ 3 }{\Lambda \plk } \left(\frac{b^3}{3}+(k-c^2) b\right)\bigg)}.
\ee
In the $a^2$ representation, we find a straightforward modification of the Wheeler-DeWit equation, 
leading (with simple adaptations from~\cite{CSHHV}) to the torsionful version of the Hartle-Hawking wave function\footnote{Note that there is a typo in the definition of $z$ in \cite{CSHHV}.} :
\be\label{HHc1}
\psi_{HH}(b,c)= {\cal N}' {\rm Ai} (-z),
\ee
with 
\be\label{zexp}
z=-\left(\frac{3 }{\Lambda \plk }\right)^{2/3}
\left(k-c^2-\frac{\Lambda a^2}{3}\right).
\ee
%[Or the Vilenkin... different contours. Let's keep the contour real.]
Using standard results in the theory of Airy functions we have:
\be\label{NandNp}
{\cal N}'=  {\cal N}\sqrt{\frac{2\pi}{\plk}}\left(\frac{\Lambda \plk}{3}\right)^{1/3}.
\ee

As in~\cite{CSHHV} we may argue that there is only one state (once the matter of the domains of variation is fixed),
taking the form of the Chern-Simons or Hartle-Hawking wave functions depending on the chosen representation. The only novelty here is that 
the state is index by the torsion $c$. We denote this state $|\Psi_{HC}(c)\rangle$ and in 
a representation-free notation we may write:
\be
\hat {\cal H}(a^2,b;c)|\Psi_{HC}(c)\rangle=0.
\ee
The explicit representations result from:
\bea
\langle b|\Psi_{HC}(c)\rangle&=&\psi_{CS}(b,c)\\
\langle a^2|\Psi_{HC}(c)\rangle&=&\psi_{HH}(a^2,c)
\eea
and proving that $\psi_{CS}(b,c)$ and $\psi_{HH}(a^2,c)$ are the Fourier transform of the other is then a particular case of the insertion of a partition of identity as defined above.

These wave functions may be called ``kinematical'' with regards to the torsion, because we have not yet imposed the conditions which we know 
they must satisfy, in {\it some}  sense,  regarding the torsion. Before proposing a prescription for these conditions (Section~\ref{wavepacks}), 
we show how the kinematical wave functions clarify the important 
issue of the normalizability of the Chern-Simons state and the Hartle-Hawking wave function.

\section{Normalizability of the Chern-Simons and Hartle-Hawking states}\label{norms}

%At this point we re-examine claims that the Chern-Simons state adapted to a real theory is normalizable, because it is a pure phase [CITE Randono]. 
%It certainly is better than the generic state for the SD theory, which is not a pure phase (it includes the real part of $Y_{CS}$)
%rendering it non-invariant under large gauge transformations, CPT violating, *** ETC, AND certainly non-normalizable according to 
%any obvious measure. 

We start by examining the sense in which the ``real'' Chern-Simons state~\cite{realK}, which is a pure phase in Lorentzian signature (just like its Euclidean cousin), 
is ``delta-function''  normalizable.  The state certainly resembles a wave extending over an infinite domain (in $b$, in mini-superspace; or in $a^2$ for its dual).
Such waves, spread over an infinite domain, are regularizable by confining them into a finite box, 
but this would be equivalent to truncating the domain of $b$ (or $a^2$) and discretizing $a^2$ (or $b$), clearly a very crude expedient. 

A more palatable alternative consists of taking the kinematical torsionful wave functions and evaluate their inner product 
for generic torsion values.  Using (\ref{bmeas}) and (\ref{kodbc1}) we find, in the $b$ representation:
\bea
\langle \Psi_{HC}(c)|\Psi_{HC}(c')\rangle&=& \int db\, \psi^\star _{CS}(b,c)\psi_{CS}(b,c')\nn\\
&=&{\cal N}^2 \int db\,   \exp{\bigg(i\frac{ 3 }{\Lambda \plk } \left( (c^2-c'^{2}) b\right)\bigg)}\nn\\
&=&2\pi\frac{\Lambda\plk}{3}{\cal N}^2 \delta(c^2-c'^2)
\eea
(note how the cubic term in $b$ in the phase cancels, leaving the right factors for a delta function).
By choosing:
\be\label{calN}
{\cal N}=\sqrt \frac{3}{2\pi \Lambda\plk }
\ee
we therefore have:
\be\label{deltanorm}
\langle \Psi_{HC}(c)|\Psi_{HC}(c')\rangle=\delta(c^2-c'^2)
\ee
giving a clear implementation of the statement that  the real Chern-Simons wave functions are ``delta-normalizable''.

We now understand why even the non-pathological real  Chern-Simons state, which is a pure phase, has infinite norm:
\be
\int db\, |\psi_{CS}(b,c)|^2
={\cal N} \int db
=\infty 
\ee
for any torsion $c$, and in particular for $c=0$. It is because 
\be
\int db\, |\psi_{CS}(b,c)|^2= \langle \Psi_{HC}(c)|\Psi_{HC}(c)\rangle=\delta(0)=\infty.\nn
\ee
In particular if we force the torsion to be zero by construction (as in the usual approach to the second
class constraints associated with it), we have:
\be\label{notorinfnorm}
%\int db\, |\psi_{CS}(b,c\equiv 0)|^2 = 
\langle \Psi_{HC}(c\equiv 0) |  \Psi_{HC}(c\equiv 0)\rangle=\infty.
\ee
By building wave packets in $c$ (possibly centred at $c=0$, but with a width), as we shall do in the next Section, the 
norm becomes finite, in perfect analogy with delta-normalizable plane waves and their wave packets. 

Having gone this far, we note that norms do not depend on the representation, so we can translate everything 
we said so far into the metric representation, and see what it implies for the Hartle-Hawking wave function. It is known that the (torsion-free) Hartle-Hawking
state is not normalizable:
\be
\int da^2\,
%\frac{da^2}{ \sqrt{2\pi}\plk}
|\psi_{HH}|^2=
{\cal N}'^2
\int da^2\,
%\frac{da^2}{ \sqrt{2\pi}\plk}
{\rm Ai}^2(z)=\infty ,
\ee
a matter which has led to several probability interpretations(see~\cite{Vilenkin,vil0} and also\cite{Halliwell} and references therein). 
Without discussing their merits, we note that this non-normalizability 
is just an expression of (\ref{notorinfnorm}) in the metric representation,
and results from fixing the torsion to zero too abruptly and rigidly, just as was found in the connection representation. 
And likewise, the problem can be removed
by introducing the torsionful Hartle-Hawking wave functions and note that these are already delta-function normalized in the sense of
(\ref{deltanorm}) (since the norm does not depend on the representation). Finite norm versions of the the Hartle-Hawking wave function can then
be found constructing wave packets in the torsion out of these delta-normalized torsionful waves.

Just to make sure, we double check our statement with a direct calculation. Bearing in mind the identity:
\be
\int  dz {\rm Ai} (z+x) {\rm Ai}(z+y)=\delta(x-y),
\ee
we can evaluate (\ref{deltanorm})  in the  $a^2$ representation as:
\bea
\langle \Psi_{HC}(c)|\Psi_{HC}(c')\rangle
&=& \int da^2 \psi^\star _{HH}(a^2,c)\psi_{HH}(a^2,c')\nn\\
&=&\frac{3}{\Lambda} \left(\frac{\Lambda\plk}{3}\right)^{4/3} {\cal N}'^2\delta(c^2-c'^2).
\eea
Hence we recover (\ref{deltanorm}) if:
\be
{\cal N}'=\sqrt{\frac{\Lambda}{3}}\left(\frac{3}{\Lambda\plk}\right)^{2/3}
\ee
which could have been found to have this value directly from (\ref{NandNp}) and (\ref{calN}). 

%As with the CS state we can build finite norm versions of the the Hartle-Hawking wave function by building wave packets 
%in torsion space from the torsionful versions of the Hartle-Hawking wave function, as we will now do. 

\section{Wave packets and the torsion condition}\label{wavepacks}
We can now consider general normalizable solutions to the Hamiltonian constraint. The $|\Psi_{HC}(c)\rangle$ 
can be seen as monochromatic waves (and so have infinite norm), but by linearly superposing them into wave packets  one 
arrives at normalizable solutions, since they are delta-function normalized. Such packets solve the Hamiltonian constraint because the Hamiltonian is linear {\it and}
does not contain $p_c$ (so, multiplying each $|\Psi_{HC}(c)\rangle$ by an amplitude, $A(c)$, dependent on $c$, still produces a
solution). 

Given the form of the delta-function normalization (cf. Eq.~(\ref{deltanorm})), the most convenient measure with which to label the
superposition is $dc^2$.  Thus, we arrive at:
\be\label{wavepackets}
|\phi\rangle =\int dc^2 A(c^2) |\Psi_{HC}(c)\rangle.
\ee
The normalization condition 
$\langle \phi|\phi\rangle=1$ becomes the statement:
\be\label{normA}
\int dc^2 |A(c^2)|^2=1,
\ee 
so, any suitably chosen $A(c^2)$ leads to well defined probabilities. These packets may 
be written in the $a^2$ or $b$ representation, applying to them (\ref{a2rep}) or (\ref{brep}), i.e.:
\bea
\phi(a^2)&=&\int dc^2 A(c^2)\psi_{HH}(a^2,c)\label{packeta2}\\
\phi(b)&=&\int dc^2 A(c^2)\psi_{CS}(b,c)\label{packetb}.
\eea
The amplitude  $A(c^2)$ is the same in both representations.

We are finally ready to construct states compliant with a quantum version of the torsion-free condition. 
They should be coherent states centred at $c=0$. Given that $c$ does not have a momentum there is some
arbitrariness in the construction. Specifically, in order to fit the measure $dc^2$, we could have replaced (\ref{Sc}) by\footnote{Note that this choice implies that the term $c^{2}$ in the Hamiltonian density is now linear in the configuration variable which may alternatively be regarded as a term proportional to the momentum \emph{of} $p_{c^{2}}$. Models containing such contributions can be candidates for emergent clocks in quantum gravity \cite{Gielen:2020abd} but we will not explore this further in the present paper.}:
\be
%\label{Sc}
%S\rightarrow S-  6\kappa V_c \int dt (\dot c p_c - \lambda  p_c)
S\rightarrow S+ \int dt ( \dot{c^2} p_{c^2} - \lambda  p_{c^2})
\ee
suggesting we build coherent states from:
\be
[\hat c^2,\hat p_{c^2}]=i\hbar\implies \hat{p}_{c^2}=-i\hbar \frac{\partial}{\partial c^2}.
\ee

As with the case for free-particles (lacking a potential term in the Hamiltonian), the fact that $p_c$ does not appear in the Hamiltonian means there is a free
length scale, $\ell$, required to define dimensionless quadratures and ``annihilation'' operators (see~\cite{freecoh}, for a 
very pedagogical review). Thus, coherent states will be eigenstates of:
\be\label{hatZ}
\hat Z=\frac{\ell^2}{ \sqrt 2}\hat{c^2}+i\frac{1}{\hbar\ell^2 \sqrt{2}}\hat{p}_{c^2}
=\frac{\ell^2}{ \sqrt 2}c^2+\frac{1}{\ell^2 \sqrt{2}} \frac{\partial}{\partial c^2}.
\ee
The scale $\ell$ can be the Planck length, Lambda's length scale  $|\Lambda|^{-1/2}$, or any function thereof
(we could also appeal to $V_c$).  In particular, we can investigate:
\be
\ell=\left(l_P^n\Lambda^{-m/2}\right)^{\frac{1}{n+m}}.
\ee
The case $n=2$ and $m=1$ is studied in~\cite{RandoMeso,Barrowhbar} and leads to interesting mesoscopic 
implications. The choice $n=1$ and $m=2$ leads to effects on the scale of $10^2$ Hz, of possible interest for gravity wave
detection. The fact that $\ell$ is not fixed is not surprising and we will comment on it at the end of this section, once
coherent states are constructed. 

The torsion-free condition may be implemented by constructing coherent states centred at the origin, i.e. by solving
 $\hat Z A(c^2)=0$, with $\hat Z$ as in (\ref{hatZ}).
These have the form:
\be\label{Amp}
A(c^2)=\frac{\exp{\left(-\frac{c^4}{4\sigma^2_{c}}\right)}}
{(2\pi \sigma_c^2)^{1/4}},
\ee
the denominator chosen so as to enforce  (\ref{normA}). Here we use the (slightly misleading) notation:
\be
\sigma_c^2=\langle c^4\rangle=2/\ell^4
\ee
to avoid the heavy alternative  $\sigma^2_{c^2}$. The amplitude  (\ref{Amp}) will be used for the rest fo this paper.

We close by noting that we could have built coherent states centred around any point $\{c_0^2,p_{c^2_0}\}$. 
For example, for the quasi-Euler theory~\cite{alex1,alex2,MZ} we should set $p_{c^2_0}=0$, but $c_0^2$ could be any constant. 
We could also
have considered squeezed coherent states. The more general expression is: 
\be\label{cohgen}
A(c^2)=\frac{1}{(2\pi \zeta^2 \sigma_c^2)^{1/4}}
\exp{\left(-\frac{(c^2-c_0^2)^2}{4\zeta^2\sigma^2_{c}}+\frac{i}{\hbar}p_{c^2_0} c^2\right)},
\ee
where $\zeta$ is the squeezing parameter.

The uncertainty in the length scale $\ell$ (and so in $\sigma_c$)
is therefore an uncertainty in the definition of squeezing, since only $\sigma_c\zeta$ appears in the states. 
Its origin lies in the fact that the conjugate momentum to $c^2$ does not appear anywhere, so it can be defined in a number of ways. 
We can saturate the Heisenberg bound imposed by the complementarity 
of $c^2$ and its momentum, but have no way to define ``balanced'' uncertainties between conjugates for an coherent state
(or how unbalanced they are for a squeezed state). 
Note that a similar situation happens for a free particle~\cite{freecoh}, amounting to the introduction of a length scale to define
dimensionless  quadratures.

%[TBD: nondimensionalization of the problem, to find out in terms of quadratures what is a balanced saturation of the 
%inequality. Squeezed states could also be defined here. Maybe what is squeezed what is not is undefined?]

%We could also consider squeezed coherent states, i.e. states that saturate the Heisenberg bound, 
%but with unequal uncertainties in the complementary variables $c^2$ and $p_{c^2}$. These would be the same as 
%(\ref{cohgen}), but with a generic $\sigma_c^2$, encoding the squeezing parameter. But this is undefined in our construction.
%Indeed, what we find is that squeezing is undefined. The reason is that the momentum conjugate to $c^2$ does not
%appear anywhere, so 

\section{The Hartle-Hawking beam}\label{secHHbeam}
%Explicit torsion-regularized wave packets}\label{HHbeam}
To complete our construction we evaluate explicit forms for the wave packets.  
It is easier to put it all together in the $b$ representation. Inserting (\ref{Amp}) into (\ref{packetb}) (recalling (\ref{kodbc1}))
leads to:
\bea
\phi(b)&=& {\cal N} (8\pi\sigma_c^2)^{1/4}\exp{\left[-\frac{9 b^2\sigma_c^2}{\Lambda^2\plk^2}
+ i\frac{ 9 }{\Lambda \plk } \left(b^3+3 k  b\right)\right]}
\nn\\
&=& (8\pi\sigma_c^2)^{1/4}\exp{\left[-\frac{9 b^2\sigma_c^2}{\Lambda^2\plk^2}\right]}\psi_{CS}(b,0).\label{CSbeam}
\eea
The last expression relates the infinite norm zero-torsion Chern-Simons state to the
finite norm wave packet built around zero-torsion. We see that it is dressed by a Gaussian, which regularizes it.
Indeed the associated  probability density is:
\be
P(b)=|\phi(b)|^2={\cal N}^2\sqrt{8\pi\sigma_c^2}\exp{\left[-\frac{18 b^2\sigma_c^2}{\Lambda^2\plk^2}\right]}.
\ee
This is just a Gaussian distribution in $b$, with variance:
\be
\sigma_b=\frac{\Lambda\plk}{6\sigma_c}.
\ee
By using (\ref{calN}) we can check that the distribution is properly normalized. To illustrate the effect of different values of $\sigma_{c}$ we can introduce the following dimensionless variables:
\begin{align}
\tilde{b} &=  \frac{3}{\beta \Lambda}\frac{b}{\plk}  \\
\tilde{\sigma}_{c} &= \beta \sigma_{c}
\end{align}
where
\begin{align}
\beta &= \bigg(\frac{3}{\Lambda \plk}\bigg)^{2/3}
\end{align}
such that
\begin{align}
P(\tilde{b}) &=  \sqrt{\frac{2}{\pi}} \tilde{\sigma}_{c}e^{-2\tilde{\sigma}_{c}^{2}\tilde{b}^{2}}
\end{align}
satisfies $\int P(\tilde{b})\tilde{b}$ = 1. 
\begin{figure}
	%[h]
	\center
	\epsfig{file=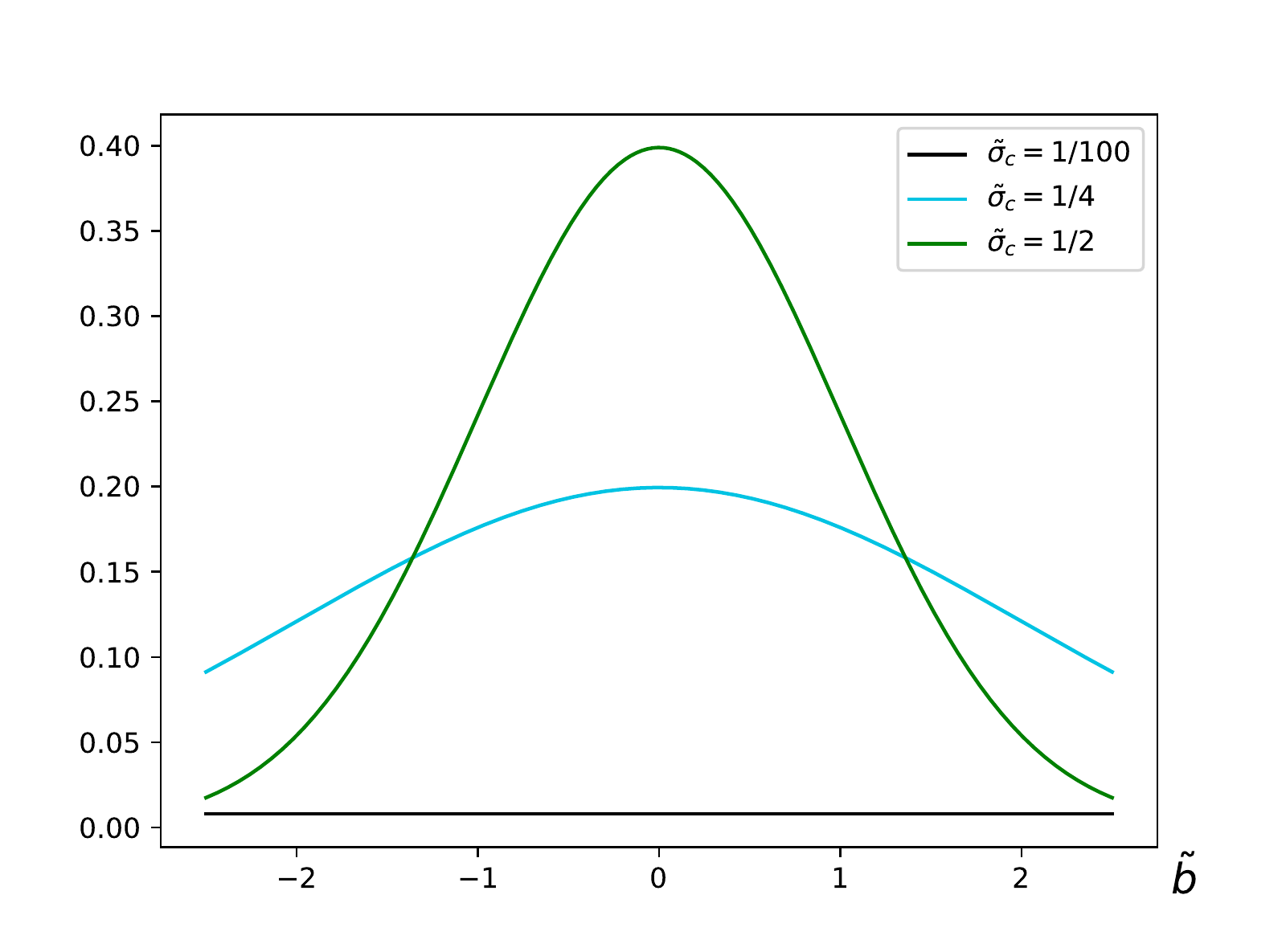,width=9.cm}
	\caption{Comparison of $P(\tilde{b})$ for various values of $\tilde{\sigma}_{c}$. }
	\label{FigureB}
\end{figure}
In Fig.~\ref{FigureB} we have plotted this distribution. As $\tilde{\sigma}_{c}$ goes to zero, the probability approaches a uniform distribution with 
infinitesimal density. The higher the fluctuations in the torsion $c$, the more peaked the distribution of $b$ around zero. Quantum fluctuations in the 
torsion seem to be acting as a ``filter'' suppressing high quantum spacetime curvature. 

\begin{figure}
%[h]
	\center
	\epsfig{file=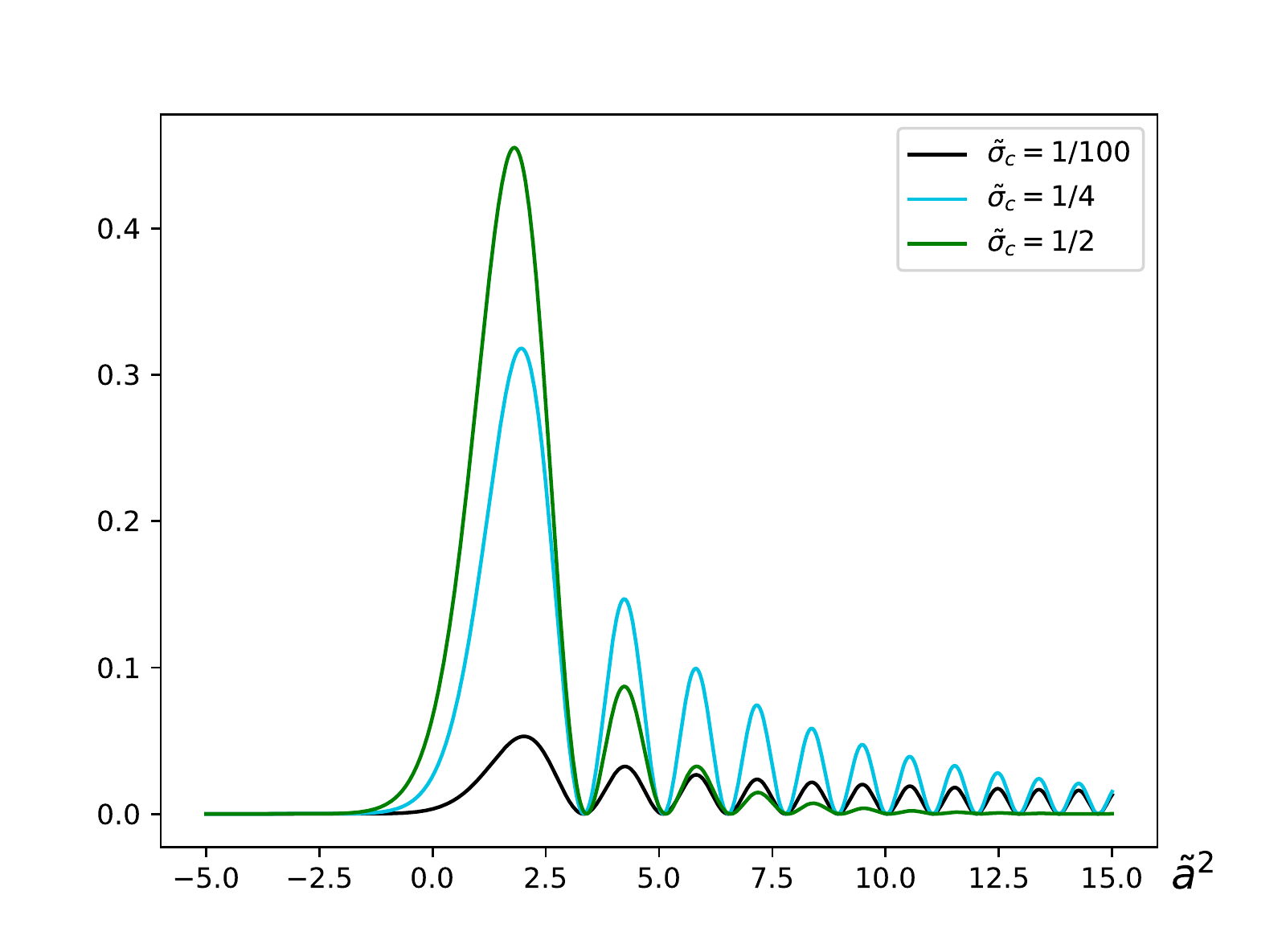,width=9.cm}
	\caption{Comparison of $|\phi(\tilde{a})|^{2}$ for various values of $\tilde{\sigma}_{c}$. For ease of illustration, plots are normalized so that $\int |\phi|^{2}d\tilde{a}^{2}=1$. Formally, as $\tilde{\sigma}_{c}\rightarrow 0$, the solution $\phi$ approaches proportionality with the Hartle-Hawking wave function $\psi_{HH}$.}
	\label{Figure2}
\end{figure}

In the $a^2$ representation we must perform the integral (\ref{packeta2}), with (\ref{Amp}). 
This is best evaluated appealing to the concept of Airy transform~\cite{Airybook}, which can
be written as:
\begin{align}
	\phi_{\alpha}(y) &= \frac{1}{\alpha}\int^{\infty}_{-\infty}f(x){\rm Ai}\bigg(\frac{y-x}{\alpha}\bigg) dx  \quad (\alpha > 0).
\end{align}
It is a standard result that the Airy transform of the Gaussian function:
\begin{align}
	{\cal G}(x) &= \frac{1}{\sqrt{\pi}}e^{-x^{2}}
\end{align}
is:
\begin{align}
\phi_{({\cal G})\alpha}(x) &= \frac{1}{|\alpha|}e^{\frac{1}{4\alpha^{3}}(x+\frac{1}{24\alpha^{3}})}{\rm Ai}\bigg(\frac{x}{\alpha}+\frac{1}{16\alpha^{4}}\bigg).
\end{align}
With this result in hand we find, after some algebra,  that our wave-packet can be written in the form
\begin{align}
\phi(\tilde{a}^{2}) &=\tilde{\cal N}	e^{\tilde{\sigma}_{c}^{2}\big(\tilde{k}-\tilde{a}^{2} + \frac{2}{3}  \tilde{\sigma}_{c}^{4}\big)}\mathrm{Ai}\bigg(\tilde{k}-\tilde{a}^{2}+  \tilde{\sigma}_{c}^{4}\bigg)
\end{align}
where we have defined the following dimensionless variables:
\bea
\tilde{a}^{2} &=&  \beta\frac{\Lambda a^{2}}{3}\nn\\
\tilde{k} &=& \beta k\nn\\
\tilde{\sigma}_{c} &=& \beta\sigma_{c}\nn\\
\tilde{\cal N}& = &(8\pi\sigma_c^2)^{1/4} {\cal N}'\nn.
%\frac{2\sqrt{\pi}\sqrt{\sigma_{c}}}{(2\pi )^{1/4}}.
\eea
This can be more suggestively written as:
\be\label{HHbeam}
\phi(a^2)=Ce^{-\lambda_{a} a^2}\psi_{HH}(a^2,c_0^2)
\ee
which we call the Hartle-Hawking beam, and illustrate in Fig.~\ref{Figure2} for various values of $\tilde{\sigma}_{c}$. 

This beam is a Hartle-Hawking wave function dressed with an exponential that makes it fall off 
at large $a^2$ faster than the usual (non-normalizable) power-law. Unlike the equivalent result for the $b$ representation (cf. Eq.~(\ref{CSbeam})),
the state being dressed has an effective torsion, formally given by:
\be
c_0^2= - \left(\frac{3}{\Lambda\plk}\right)^{2}\sigma_c^4.
\ee
This shifts the zeros of the wave function, but does not affect their spacing. The phenomenology described in~\cite{RandoMeso}, therefore,
is not expected to change. The suppression exponent, $\lambda_a$, is given by:
\be
\lambda_a=\frac{3}{\Lambda\plk}\sigma_c^2=\frac{54 V_c^2}{\Lambda l_P^4 \ell ^4}.
\ee
Depending on the torsion scale $\ell$ and on the observation volume this could have meso- and macroscopic observable effects.
We will return to this matter elsewhere.

Finally, we note that 
for $a^2<0$ the dressing exponential blows up, but the Airy function decays exponentially even faster. The Hartle-Hawking beam is a perfectly
well behaved distribution. 
For completeness we include the irrelevant expression for the proportionality constant $C$ in (\ref{HHbeam}):
\be
C=\exp{\left[\left(\frac{3}{\Lambda\plk}\right)^{2} \sigma_c^2\left(k+\frac{2}{3}\left(\frac{3}{\Lambda\plk}\right)^{2}\sigma_c^4\right)\right]}.
\ee

.

%\section{Pending issues, demonstrated in mini-superspace}
\section{Beyond mini-superspace}
%[Do it the other way round (in brackets the mini-superspace version)]

One may wonder if our normalization procedure can extend beyond mini-superspace. 
The answer is yes, as we proceed to schematically show, drawing heavily on~\cite{realK,GenHH} (to which we refer the reader for details). 
In general, we should apply to our constructions the correspondence:
\bea
a^2&\rightarrow E^a_i\\
b&\rightarrow  K^i_a
\eea
where $E^a_i$ are the densitized inverse triads, $K^i_a$ are the extrinsic curvature 1-forms (the imaginary part of the Ashtekar Self-Dual connection, $A^i_a$),
with $i$ the $SU(2)$ indices and $a$ the spatial coordinate indices. The spatial connection $\Gamma^i_a$ 
(the real part of $A^i_a$) can be split into a torsion-free part $\bar{\Gamma}^{i}$ and a 
contorsion $\tilde{\Gamma}^{i}$:
\be
\Gamma^i=\bar \Gamma^i(E)+\tilde\Gamma^i
\ee
In mini-superspace, $k=\pm 1$ provides an example of the effect of the first term, $c$ of the second. 
We shall say more about this splitting in the next Section.
%[SHIFT BACK the parity violating Cartan spiral staircase. ] [DO THEIR CURVATURES ALWAYS SPLIT?]

With this dictionary in hand we can reproduce most of the steps in this paper in a setting beyond  mini-superspace.
The Fourier transform defined in (\ref{FT})  becomes the one proposed in~\cite{GenHH}:
\be\label{FTgen}
\psi_E(E,\Gamma) =\prod_{\vec x,a,i} \int \frac{d [ K^i_a (\vec x))]}{\sqrt{2\pi l_P^2}} e^{-\frac{i}{l_P^2} E^a_i(\vec x)  K^i_a (\vec x)}\psi_K(K,\Gamma),
\ee
where the product include all spatial points $\vec x$, discretized or as a continuum of infinitesimals. 
The inner product that generalizes (\ref{bmeas}) is:
\bea
\langle \psi_1(\Gamma)|\psi_2(\Gamma')\rangle&=&\prod_{\vec x,a,i} \delta V_c  \int  d [ K^i_a (\vec x))] 
\psi_{K1}^\star(K^i_a (\vec x)), \Gamma ^i_a (\vec x))\nn\\
&& \psi_{K2} (K^i_a (\vec x)), \Gamma ^{i'}_a (\vec x))\label{geninnprod}
\eea
where $\delta V_{c}$ is the volume element around point $\vec x$ introduced for convenience in the next steps, mimicking
our construction in mini-superspace.  Indeed, 
the generalization of  (\ref{kodbc1}) is~\cite{realK}:
%\begin{widetext}
\bea
\psi_{CS}&=&{\cal N}\exp{\left[-\frac{3i}{\Lambda l_P^2}\int K^i\,  {}^{(3)}\! R^i  - \epsilon_{ijk}\frac{K^i K^j K^k}{6} \right]} 
\label{realwithreal2}.
\eea
%\end{widetext}
This is the general solution to the Hamiltonian constraint in the connection representation, with the torsion left in.
It should be seen as the general state which is still kinematical with regards to the torsion, but not otherwise.
It has dual metric representations, with formal definition and examples beyond mini-superspace given in~\cite{GenHH}.
It factorizes as:
\begin{widetext}
\bea
\psi_{CS}&=&\prod_{\vec x,a,i} {\cal N}_{ai}(\vec x)
\exp{\left[-\frac{3i\delta V_c }{\Lambda l_P^2} \epsilon^{abc} \left( K^i_a\,  {}^{(3)}\! R^i_{bc}  - \epsilon_{ijk}\frac{K^i_a K^j_b K^k_c}{6} \right) \right]} .
\eea
\end{widetext}
As in mini-superspace, when we evaluate the inner product for states with different $\Gamma^i$ (using (\ref{geninnprod})), 
the cubic term in $K^i$ cancels out, 
leaving us only with linear terms that lead to delta-functions. Hence, in general we have:
\be
\langle \psi_{CS}(\Gamma)|\psi_{CS}(\Gamma')\rangle=\prod_{\vec x,a,i} \delta({}^{(3)}\! \tilde R^{ia} (\vec{x}) -  {}^{(3)}\! \tilde R^{ia'} (\vec{x}))
\ee
with $ {}^{(3)}\!  \tilde R^{ia} = \epsilon^{abc}  {}^{(3)}\!  R^i_{bc}$ and 
\be
{\cal N}_{ai}(\vec x)=\sqrt{\frac{3}{2\pi\Lambda l_P^2}}.
\ee
We can now build wave packets out of these states with generic amplitudes $A(\Gamma)$. 

We see that a general treatment generates delta normalization most naturally in terms of the dual of the 3-curvature. 
In mini-superspace this reduces to the single parameter:
\be\label{kc}
k_c=k-c^2
\ee
Obviously if we fix $k$ this is equivalent to the normalization in $c^2$ found before. Nothing changes with regards to the 
definition of the Hartle-Hawking beam. But an intriguing possibility is raised. Given that states are now indexed by $k_c$,
could states with different $k$ but the same $k_c$ be the same? Could there be cross talk between the 3 types of FRW Universe?
We will investigate this possibility further in~\cite{QuantFlat}, with direct reference to the infamous flatness problem in Big Bang cosmology.

%. This is the term giving $c^2$ or $k-c^2$ in mini-superspace (but with Bianchi I we get
%factors of the form $\delta(c_1 c_2 - c_1' c_2')$ and cyclic permutations). This is achieved with 

%In general we can consider states of different $k$ and $c$. We find that with the norms defined the states should be indexed by:
%\be
%k_c=k-c^2
%\ee
%which is the mini-superspace version of the 3-curvature including a torsion-free part and the torsion component. 

%Thus, it seems that states with the same $k_c$ are the same state. What does this mean? For some values of $c$ there
%all 3 $k$ correspond to the same state!

\section{Torsion in the Einstein-Cartan theory}
But there is more. As already noted in Section~\ref{background}, in the canonical formalism there are two types of torsion: 
that contained in $\Gamma^i$ ($c$ in mini-superspace), and that contained in $K^{i}$ (the $T$ in $b=\dot a/N-T$). 
They are treated very differently classically: the first leads to second class constraints, the second is set to zero by a Hamilton equation. 
according to our procedure, quantum mechanically they are treated even more differently. And yet, classically one can convert one type of curvature into the
other as we now show. 

To recap, the action for the Einstein-Cartan theory is given in differential forms notation by:

\begin{align}
S[e,\omega] &= \int \eps_{IJKL}e^{I}e^{J}R^{KL}(\omega) \label{ecaction}
\end{align}
where $I,J,\dots$ are $SO(1,3)$ indices, $e^{I} = e^{I}_{\mu}dx^{\mu}$ is the co-tetrad, $\omega^{I}_{\ph{I}J} = \omega^{I}_{\ph{I}J\mu}dx^{\mu}$ is the spin connection and $R^{IJ} = d\omega^{IJ}+ \omega^{I}_{\ph{I}K}\omega^{KL}$ is the curvature two-form. Considering the action (\ref{ecaction}) to be a functional of $e^{I}$ and $\omega^{IJ}$ as independent fields, variation of $\omega^{IJ}$ yields the following classical equation of motion 

\begin{align}
de^{I}+\omega^{I}_{\ph{I}J}e^{J} \equiv T^{I} = 0 \label{torsionfree}
\end{align}
where $T^{I}$ is the torsion two-form. Alternatively, one may consider the following decomposition of $\omega^{IJ}$ at the level of the action:

\begin{align}
\omega^{IJ} &= \bar{\omega}^{IJ}(e) + C^{IJ}
\end{align}
where $\bar{\omega}^{IJ}(e)$ is the unique solution of equation (\ref{torsionfree}) for $\omega^{IJ}$ assuming that $e^{I}_{\mu}$ considered as a matrix is invertible and $C^{IJ}$ is the contorsion one-form. Using this ansatz in (\ref{ecaction}) and integrating by parts leads to the action

\begin{align}
S[e,C] &= \int \eps_{IJKL}e^{I}e^{J}\bigg(R^{KL}(e)+ C^{K}_{\ph{K}M}C^{ML} \bigg).\label{ecaction2}
\end{align}
The first term is simply the Einstein-Hilbert action of General Relativity. The action can be written as follows:

\begin{align}
S[e,C] &=  \int  (2\bar{R}+ C^{I}_{\ph{I}KI}C^{KJ}_{\ph{KJ}J}- C^{I}_{\ph{I}JK}C^{JK}_{\ph{JK}I}) \sqrt{-g}d^{4}x
\end{align}
where $C^{IJK} \equiv e^{K\mu}C^{IJ}_{\mu}$. In passing to the Hamiltonian formulation of this theory, one can consider a phase space coordinatized by the spatial metric $q_{ab}(x^{c},t)$ - where $x^{a}$ are spatial coordinates - its momentum $\Pi^{ab}(x^{c},t)$ as well as $C^{IJK}(x^{a},t)$ and its momentum ${\cal P}_{IJK}(x^{a},t)$. Equivalence of the classical Euler-Lagrange equations and Hamilton's equations would require the imposition of primary constraints ${\cal P}_{IJK}\approx 0$ and requiring the preservation of this constraint in time would impose further that $C^{IJK} \approx 0$, which is indeed the Euler-Lagrange equation for $C^{IJK}$. 

Explicitly, the Hamiltonian density $H$ will be given by a sum of constraints:

\begin{align}
H &= N{\cal H} + N^{a}{\cal H}_{a} + \lambda^{IJK}{\cal P}_{IJK} + \xi_{IJK}C^{IJK}
\end{align}
where $N$ and $N^{i}$ are the lapse and shift functions whilst $\lambda^{IJK}$ and $\xi_{IJK}$ are Lagrange multiplier fields and

\begin{align}
{\cal H} &=  -2\sqrt{q}\bar{\cal R} + \frac{2}{\sqrt{q}}\bigg(\Pi^{ab}\Pi_{ab}-\frac{1}{2}\Pi^{a}_{\ph{a}a}\Pi^{b}_{\ph{b}b}\bigg)  \nn\\
& - \sqrt{q}\bigg(C^{I}_{\ph{I}KI}C^{KJ}_{\ph{KJ}J}- C^{I}_{\ph{I}JK}C^{JK}_{\ph{JK}I}\bigg)\\
{\cal H}_{i} &= -4\sqrt{q}q_{ac}\partial_{b}\bigg(\frac{1}{\sqrt{q}}\Pi^{cb}\bigg)
\end{align}
where $q$ is the determinant of $q_{ab}$.

In this paper 
we adopted the approach of implementing \emph{only} the constraints ${\cal H}$ and ${\cal H}_{i}$ on quantum states 
and then building wave packets that represent coherent states around zero contorsion. But with an integration by parts
we have now put {\it all} the torsion into the last procedure.
To illustrate the point in mini-superspace, this is the action we would start from:
\begin{widetext}
\begin{equation}
S[a,\Pi,T,c]=6\kappa V_{c}\int dt\bigg(a^{2}\dot{\Pi}+Na\bigg(\Pi^{2} +k + T^{2}-c^2 -\frac{\Lambda }{3}%
a^{2}\bigg)\bigg). \label{frwact3} 
\end{equation}
\end{widetext}
instead of (\ref{Sg}). Thus the second type of torsion $T$ (that beforehand revealed by the dynamics) is transferred to the spatial curvature. 
We could also transfer just some of it or none at all.  In fact, the action (\ref{frwact3}) has a local (in time) $SO_{C}(1,1)$ (complexified 2d Lorentz group) gauge invariance - reflected in the invariance of $T^{2}-c^{2}$. 

Instead of (\ref{kc}) our states are now indexed by:
\be
k_c=k+T^2-c^2.
\ee
Nothing changes in our construction of wave packets, and yet the quantum theory looks totally different with now two dimensions of torsion space present in the problem. The million dollar question with regards to the other phenomenology is whether or not we have broken local Lorentz invariance
in our treatment of the torsion.

\section{Conclusions}
The main achievement of this paper was to find a regular, normalizable version of the Hartle-Hawking wave function, which we called
the Hartle-Hawking beam. The crucial point is not to impose the torsion-free condition too soon, namely classically, before
quantization. The latter may seem innocuous, but would amount to freezing quantum fluctuations in the torsion, implying 
essentially states of the form (\ref{wavepackets}) with
\be
A(c^2)=\delta(c^2)
\ee
so that their norm is:
\be
\int  dc^2 |A(c^2)|^2=\int dx \delta^2(x)=\infty. 
\ee
With the benefit of hindsight this should have been obvious. Fixing the torsion to zero is to consider a wave function with 
an amplitude in torsion space that is a delta function. The integral of the square of a delta function
is infinite. By allowing the torsion to fluctuate, even with a state of minimal uncertainty, such as a coherent state centred at the origin, 
we remove this infinity. We called such a wave packet (which is a Gauss-Airy function), the Hartle-Hawking beam. 

We probably would not have found the Hartle-Hawking beam, had we not received inspiration from its dual version, the real Chern-Simons state. 
It has been known for a while that the Euclidean version of that state is delta-function normalizable, since it is a pure phase.  
A clear statement of what that means was missing
until this paper. Furthermore, it has been found recently that the Lorentzian state is more similar to the Euclidean state than thought before, 
as long as an explicitly real theory
is used~\cite{realK}. Again, the matter relates to what to do with torsion in a complexified theory. Quite often reality conditions and torsion-free conditions
are mixed and confused. A separation of the two was required in order to find a non-pathological normalizable version of the Chern-Simons state. 
Its Fourier dual is precisely the Hartle-Hawking beam.  

Beyond these achievements, many questions remain to be answered on a mathematical level. But foremost we can look forward to investigating the phenomenology of torsion fluctuations, particularly in situations where a mesoscopic scale is present~\cite{RandoMeso,Barrowhbar}. 
A major issue to be cleared is whether or not our quantum treatment of torsion has broken local Lorentz invariance.
We will return to this matter in the not too distant future

\section*{Acknowledgements}

We would like to thank Stephon Alexander, Gabriel Herczeg and  Simone Speziale for enlightening discussions related to this paper. 
This work was supported by the STFC Consolidated Grant ST/L00044X/1 (JM) and by the Grant Agency of the Czech Republic, GA\u{C}R grant 20-28525S (TZ).

%\bibliographystyle{unsrtnat}
%\bibliography{references}

\end{document}